\documentstyle[epsfig]{article}
\input epsf.tex
\begin{document}
\bibliographystyle{prsty}
\title{Density Matrix Renormalization: A Review of the Method and its Applications}
\author{Karen Hallberg\\
\small\it Centro At\'omico Bariloche and Instituto Balseiro\\
\small\it 8400 Bariloche, Argentina \\}
\maketitle

\begin{abstract}
The Density Matrix Renormalization Group (DMRG) has become a powerful
numerical method that can be applied to
low-dimensional strongly correlated fermionic and bosonic systems. It
allows for a very precise calculation of static, dynamical and  thermodynamical
properties. Its field of applicability has now extended beyond Condensed
Matter, and is successfully used in Statistical Mechanics and High
Energy Physics as well.
In this article, we briefly review the main aspects of the
method. We also comment on some of the most relevant applications so as to
give an overview on the scope and possibilities of DMRG and mention the
most important extensions of the method such as the calculation of
dynamical properties, the application to
classical systems, inclusion of temperature, phonons and disorder, field 
theory, time-dependent properties and the {\it ab initio} calculation of 
electronic states in molecules.
\end{abstract}

\section{Introduction}

The basics of the Density Matrix Renormalization Group were developed
by S. White in 1992\cite{white1} and since then DMRG 
has proved to be a very powerful
method for low dimensional interacting systems. Its remarkable accuracy
can be seen for example in the spin-1 Heisenberg chain: for a system of
hundreds of sites a precision of $10^{-10}$ for the ground state energy 
can be achieved.
Since then it has been applied to a great variety of systems and problems
including, among others,
spin chains and ladders, fermionic and bosonic systems, disordered models, 
impurities and molecules and 2D electrons in high magnetic fields.
It has also been improved substantially in several directions like
two (and three) dimensional (2D) classical systems, stochastic models,
the presence of phonons, quantum chemistry, field theory,  
the inclusion of temperature and the calculation of dynamical and time-dependent 
properties.
Some calculations have also been performed in 2D quantum systems.
All these topics are treated in detail and in a pedagogical way in the
book \cite{book}, where the reader can find an extensive review on
DMRG. In this article we will attempt to cover the different areas where it
has been applied without entering into details but in a few cases, where 
we have chosen some representative contributions. We suggest the interested 
reader to look for further information in the referenced work. 
Our aim here is to give the reader a general overview on the
subject. 

One of the most important limitations of numerical calculations in finite
systems is the great amount of states that have to be considered and its
exponential growth with system size. Several methods have been introduced in
order to reduce the size of the Hilbert space to be able to reach larger
systems, such as Monte Carlo, renormalization group (RG) and DMRG. Each
method considers a particular criterion of keeping the relevant information.

The DMRG was originally developed to overcome the problems that arise in
interacting systems in 1D when standard RG procedures
were applied. 
Consider a block B (a block is a collection of sites) where
the Hamiltonian $H_B$ and end-operators are defined.
These traditional methods consist in putting together
two or more blocks (e.g. B-B', which we will call the superblock),
connected using end-operators, in a
basis that is a direct product of the basis of each block, forming
$H_{BB'}$. This Hamiltonian is then diagonalized, the
superblock is replaced by a new effective block  $B_{new}$ formed by a
certain number $m$ of
lowest-lying eigenstates of $H_{BB'}$ and the iteration is continued (see
Ref.~\cite{white2}). 
Although it has been used successfully in certain cases,
this procedure, or similar versions of it, has been applied to several
interacting systems with poor performance. For example, it has been
applied to the 1D Hubbard model keeping $m\simeq 1000$ states. For 16
sites, an error of 5-10\% was obtained \cite{braychui}. Other
results\cite{panchen} were also discouraging. A better performance was
obtained \cite{xiangghering} by adding a single site at a time rather
than doubling the block size.
However, there is one case where a similar version of this  method applies
very well: the Kondo (and Anderson) model. 
Wilson\cite{wilson} mapped the one-impurity
problem onto a one-dimensional lattice with exponentially descreasing
hoppings. The difference with the method explained above is that in
this case, one site (equivalent to an ``onion shell") is added at each
step and, due to the exponential decrease of the hopping, very accurate
results can be obtained.

Returning to the problem of putting several blocks together, the main
source of error comes from the election of eigenstates of
$H_{BB'}$ as representative states of a superblock. Since $H_{BB'}$ has no
connection to the rest of the lattice, its eigenstates may have unwanted
features (like nodes) at the ends of the block and this can't be improved
by increasing the number of states kept. Based on this consideration,
Noack and White\cite{noackwhite} tried including different boundary
conditions and boundary strengths. This turned out to work well for single
particle and Anderson localization problems but, however, it did not
improve the results significantly for interacting systems.
These considerations led to the idea of taking a larger superblock that
includes the blocks $BB'$, diagonalize the
Hamiltonian in this large superblock and then somehow project the most
favorable states onto $BB'$. Then $BB'$ is replaced by $B_{new}$. 
In this way, awkward features in the boundary
would vanish and a better representation of the states in the infinite
system would be achieved. White\cite{white1,white2} proposed the
density matrix as the
optimal way of projecting the best states onto part of the system and this
will be discussed in the next section. Some considerations concerning the effect 
of boundary conditions in the nature of the states kept (and an analogy to the 
physics of black holes) is given in \cite{gaite}. 
The justification of using the density matrix is given in detail in
Ref.\cite{book}. 
A very easy and pedagogical way of understanding the basic functioning of
DMRG is applying it to the calculation of simple quantum problems like one
particle in a tight binding chain \cite{whitebook,sierraparticle}.

In the following Section we will briefly describe the standard method; in
Sect. 3 we will mention some of the most important applications; in Sect.
4 we review the most relevant extensions to the method and
finally in Sect. 5 we concentrate on the way dynamical calculations can be  
performed within DMRG. 

\section{The Method}

The DMRG allows for a systematic truncation of the Hilbert space 
by keeping the most probable states describing a wave function
({\it e.g.~}the ground state)
instead of the lowest energy states usually kept in
previous real space renormalization techniques.

\begin{figure}
\begin{center}
\epsfxsize=2.0in
\epsfysize=1.0in
\epsffile{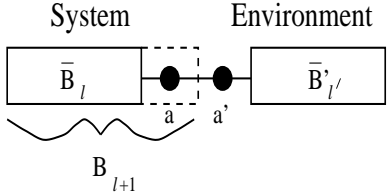}
\caption[]{A scheme of the superblock (universe) configuration for the
DMRG algorithm\cite{white2}.}
\end{center}
\label{figSuperblock}
\end{figure}

The basic idea consists in starting from a small system ({\it e.g} with
$N$ sites) and then gradually increase its size (to $N+2$, $N+4$,...)
until the desired length is reached.
Let us call the collection of $N$ sites the {\it universe} and divide it
into two parts: the {\it system} and the {\it environment}
(see Fig.\ \ref{figSuperblock}). 
The Hamiltonian is constructed in the {\it universe} and its ground state
$|\psi_0>$ is obtained. This is considered as the state of the {\it
universe} and called the {\it target state}. It has components on the {\it
system} and the {\it environment}. 
We want to obtain the most relevant states of the {\it system}, i.e.,
the states of the {\it system} that have largest weight in
$|\psi_0\rangle$.
To obtain this, the {\it environment} is considered as a statistical bath
and the density matrix\cite{feynman} is used to obtain the desired
information on the {\it system}. So instead of keeping eigenstates of the
Hamiltonian in the block ({\it system}),
we keep eigenstates of the density matrix.
We will be more explicit below.


Let's define block [{\bf B}] as a finite chain with $l$ sites, having an
associated Hilbert space with, $m$ states
where operators are defined (in particular the
Hamiltonian in this finite chain, $H_B$ and the operators
at the ends of the block, useful for linking it to other chains or
added sites). Except for the first iteration,
the basis in this block isn't explicitly known due to previous basis
rotations and reductions. The
operators in this basis are matrices and the basis states are
characterized by quantum numbers (like $S^z$, charge or number of
particles, etc).
We also define an added block or site as [{\bf a}] having $n$ states.
A general iteration of the method is described below:   

i) Define the Hamiltonian $H_{BB'}$ for the superblock (the {\it
universe})
formed by putting together two blocks [{\bf B}] and [{\bf B'}]
 and two added sites [{\bf a}] and [{\bf a'}] in this way:
[{\bf B a a' B' }]
(the primes are only to indicate additional blocks, but the primed
blocks have the same structure as the non-primed ones; this can vary, see
the finite-size algorithm below). In general, blocks [{\bf B}] and
[{\bf B'}] come from the previous iteration. The total Hilbert space of
this  superblock is the direct product of the individual spaces
corresponding to each block and the added sites. In practice a quantum
number of the superblock can be fixed (in a spin chain for example one can
look at the total $S^z=0$ subspace), so the total number of states in the
superblock is much smaller than $(mn)^2$. In some cases, as the quantum number
of the superblock consists of the sum of the quantum numbers of the individual
blocks, each block must contain several subspaces (several
values of $S^z$ for example). 
Here periodic boundary conditions can be attached to the ends and a
different block layout should be considered (e.g. [{\bf B a B' a' }])
to avoid connecting blocks [{\bf B}] and [{\bf B'}] which takes longer to
converge. The boundary conditions are between [{\bf a'}] and
[{\bf B}]. For closed chains the performance is poorer than for open boundary 
conditions \cite{white2,scalapino}.

ii) Diagonalize the Hamiltonian $H_{BB'}$ to obtain the ground state
$|\psi_0\rangle$ (target state)
using Lanczos\cite{lanczos} or Davidson\cite{davidson} 
algorithms. Other states could also be kept, such as the first excited ones: they are all
called {\it target states}.

iii) Construct the density matrix: 
\begin{equation}
\rho_{ii'}=\sum_j \psi_{0,ij}\psi_{0,i'j} 
\end{equation}
on block [{\bf B a}], where $\psi_{0,ij}=\langle i\otimes j|\psi_0\rangle
$, the states $|i\rangle $ belonging to the
Hilbert space of the block [{\bf B a}] and the states $|j\rangle $
to the block [{\bf B' a'}].
The density matrix considers the part [{\bf B a}] as a system and 
[{\bf B' a'}], as a statistical bath.
The eigenstates of $\rho$ with the  highest
eigenvalues correspond to the most probable states (or equivalently
the states with highest weight) of block [{\bf B a}] in the ground state
of the whole superblock.
These states are kept up to a certain cutoff, keeping a total of
$m$ states per block. The density matrix eigenvalues sum up to unity and
the truncation error, defined as the sum of the density matrix
eigenvalues corresponding to discarded eigenvectors, gives a
qualitative indication of the accuracy of the calculation.

iv) With these $m$ states a rectangular matrix $O$ is formed and it is
used to change basis and reduce all operators defined in  
[{\bf B a}]. This block [{\bf B a}] is then renamed as block [{\bf
B$_{new}$}] or simply  [{\bf B}] (for example, the Hamiltonian in block 
[{\bf B a}], $H_{Ba}$, is transformed into $H_{B}$ as 
 $H_{B}=O^\dagger H_{Ba} O$).

v) A new block [{\bf a}] is added (one site in our case) and the new
superblock [{\bf B a a' B'}] is formed as the direct product of the
states of all the blocks.

vi) This iteration continues until the desired length is achieved. At  
each step the length is $N=2l+2$ (if [{\bf a}] consists of one site).

When more than one target state is used, {\it i.e} more than one state
is wished to be well described, the density matrix is defined as:
\begin{equation}
\label{eq:pl}
\rho_{ii'}=\sum_l p_l \sum_j \phi_{l,ij} \phi_{l,i'j}
\end{equation}
where $p_l$ defines the probability of finding the system in the target
state $|\phi_l\rangle $ (not necessarily eigenstates of the Hamiltonian).

The method described above is usually called the 
{\it infinite-system algorithm}
since the system size increases at each iteration. There is a way to increase
precision at each length $N$ called the {\it finite-system algorithm}.
It consists of fixing the lattice size and zipping  a couple of
times until convergence is reached.  In this case and for the
block configuration [{\bf B a a' B' }], $N=l+1+1+l'$ where $l$ and
$l'$  are the number of sites in $B$ and $B'$ respectively. In this step 
the density matrix is used to project onto the left $l+1$ sites. In order
to keep $N$ fixed, in the next block configuration, the right block $B'$
should be defined in $l-1$ sites such that $N=(l+1)+1+1+(l-1)'$. The
operators in this smaller block should be kept from previous iterations
(in some cases from the iterations for the system size with
$N-2$)\cite{book}.

The calculation of static properties like correlation functions is easily
done by keeping the operators in question at each step and performing the
corresponding basis change and reduction, in a similar manner as done with
the Hamiltonian in each block\cite{white2}. The energy and measurements
are calculated in the superblock. In Ref.\cite{osborne} an interpretation of 
the correlation functions of systems at criticality
 is given in terms of wave function entanglement, conjecturing a modification of
DMRG for these cases that preserves the entanglement.

A faster convergence of Lanczos or Davidson algorithm is achieved by
choosing a good trial vector\cite{cavo,white4}.
An interesting analysis on DMRG accuracy is done in Ref. \cite{ors}.
Fixed points of the DMRG and their relation to matrix product wave
functions were studied in \cite{ostlund} and an analytic formulation
combining the block renormalization group with variational and Fokker-Planck
methods in \cite{delgadorg}.  The connection of the method with quantum
groups and conformal field theory is treated in \cite{sierraqg}.
There are also interesting connections between
the density matrix spectra and integrable models\cite{peschelint} via corner
transfer matrices. These articles give a deep insight into the essence of the
DMRG method.

\section{Applications}

Since its development, the number of papers using DMRG has grown
enormously and other improvements to the method have been performed.
We would like to mention some applications where this method has
proved to be useful. Other applications related to further developments of 
the DMRG will be mentioned in Sect. 4.


A very impressive result with unprecedented accuracy was obtained by
White and Huse \cite{white3} when calculating the spin gap in a $S=1$
Heisenberg chain obtaining $\Delta=0.41050 J$.  They also calculated
very accurate spin correlation functions and excitation energies for
one and several magnon states and performed a very detailed analysis of
the excitations for different momenta.  They obtained a spin correlation
length of 6.03 lattice spacings.  Simultaneously S{\o}rensen and
Affleck\cite{sorensen} also calculated the structure factor and spin
gap for this system up to length 100 with very high accuracy,
comparing their results with the nonlinear $\sigma$ model. In a 
subsequent paper\cite{sorensen2} they applied the DMRG to the
anisotropic $S=1$ chain, obtaining the values for the Haldane gap. They  
also performed a detailed study of the $S=1/2$ end excitations in an
open chain. Thermodynamical properties in open $S=1$ chains such as specific 
heat, electron paramagnetic resonance (EPR) and magnetic susceptibility 
calculated using DMRG gave an excellent fit to experimental data, confirming 
the existence of free spins 1/2 at the boundaries\cite{batista}.
A related problem, {\it i.e.} the effect of non-magnetic impurities in
spin systems (dimerized, ladders and 2D) was studied in \cite{laukamp,ng}.
In addition, the study of magnon interactions and magnetization of $S=1$ 
chains was done in \cite{affleck3}, supersymmetric spin chains 
modelling plateau transitions in the integer quantum Hall effect in 
\cite{marston4} and ESR studies in these systems was considered in 
\cite{sieling}. 
For larger integer spins there have also been some studies. Nishiyama 
and coworkers\cite{nishi} calculated the low energy spectrum and
correlation
functions of the $S=2$ antiferromagnetic Heisenberg open chain. They
found $S=1$ end excitations (in agreement with the Valence Bond Theory).
Edge excitations for other values of $S$ have been studied in
Ref. \cite{qinedge}.
Almost at the same time Schollw{\"o}ck and  
Jolicoeur\cite{uli1} calculated the spin gap in the same system, up to
350 sites, ($\Delta=0.085 J$), correlation functions that showed
topological order and a spin correlation length of 49 lattice
spacings.  More recent accurate studies of $S=2$ chains are found in
\cite{wangqin,wada,capone} and of $S=1$ chains in staggered magnetic fields
\cite{yulu2} 
including a detalied comparison to the non-linear sigma model in \cite{ercolesi}. 
In Ref. \cite{qin} the dispersion of the single magnon band and
other properties of the $S=2$ antiferromagnetic Heisenberg chains were
calculated.

Concerning $S=1/2$ systems, DMRG has been crucial for obtaining the   
logarithmic corrections to the $1/r$ dependence of the spin-spin
correlation functions in the isotropic Heisenberg model
\cite{karen12}. For this, very accurate values for the energy and
correlation functions were needed. 
For $N=100$ sites an error of $10^{-5}$ was 
achieved keeping $m=150$ states per block, comparing with the exact
finite-size Bethe Ansatz results.
For this model it was found that the data for the correlation
function has a very accurate scaling behaviour and advantage of
this was taken
to obtain the logarithmic corrections in the thermodynamic limit.
Other calculations of the spin correlations have been performed for
the isotropic \cite{boos,shiroishi} and anisotropic cases \cite{hiki}. 
Luttinger liquid behaviour with 
magnetic fields have been studied in \cite{luttinger}, field-induced gaps in
\cite{lou2}, anisotropic 
systems in \cite{caprara,hieida} and the Heisenberg model with a weak link 
in \cite{byrnes}. An analysis of quantum critical points and critical 
behaviour in spin chains by combining DMRG with
finite-size scaling was done in \cite{marston2}.

Similar calculations have been performed for the $S=3/2$ Heisenberg
chain \cite{karen32}. In this case a stronger 
logarithmic correction to the spin correlation function was found.
For this 
model there was interest in obtaining the central charge $c$ to 
elucidate whether this model corresponds to the same universality
class as the $S=1/2$ case, where the central charge can be obtained from 
the finite-size scaling of the energy. 
Although there have been previous attempts\cite{adriana}, these
calculations presented difficulties since they involved also a term $\sim
1/\ln^3N$. With the DMRG the value $c=1$ was clearly obtained.

In Ref. \cite{yamashita}, DMRG was applied to an effective spin
Hamiltonian obtained from an SU(4) spin-orbit critical state in 1D.
Other applications were done to enlarged symmetry cases with SU(4) symmetry
in order to study coherence in arrays of quantum dots\cite{onu}, to 
obtain the phase diagram for 1D spin orbital models\cite{affleck2} and dynamical 
properties in a magnetic field\cite{haas2}.

Dimerization and frustration have been considered
in Refs. \cite{bursill,dim1,dim2,dim3,dim4,dim5,dim6,kaburagi,hikihara2} and
alternating spin chains in \cite{patialt}.

The case of several coupled spin chains
(ladder models) have been investigated in \cite{noackchain,kawaguchi,trumper,
roger,hikihara3}, spin ladders with cyclic four-spin exchanges in 
\cite{schmid,hikihara, nunner,honda} and Kagome antiferromagnets in 
\cite{patising}. 
Zigzag spin chains have been considered in \cite{maeshima,itoi,
yulu} and spin chains of coupled triangles in \cite{triang,schotte,gendiar3}.
As the DMRG's performance is optimal in open systems, an interesting 
analysis of the boundary effect on correlation functions is done in 
\cite{scalapino}. 
Magnetization properties and plateaus for quantum spin ladder
systems\cite{tandon,lou,wanglu} have also been studied. 
An interesting review on the applications to  some exact and analytical 
techniques 
for quantum magnetism in low dimension, including DMRG, is presented in 
\cite{patisen}.

There has been a great amount of applications to fermionic systems
such as 1D  Hubbard and t-J models 
\cite{noackhubb,penc,bulut,malvezzi,eric,zhang2,
baeriswyl,aoki,daul2,qin2,armando}, Luttinger liquids with boundaries\cite{kurt},
the Falicov-Kimball model \cite{falicov}, the quasiperiodic Aubry-Andre 
chain\cite{schuster} and Fibonacci-Hubbard models\cite{fibonacci}. It has also 
been applied  to field theory\cite{gaite,phi4}.
The method has been very successful for several band Hubbard 
models\cite{srinivasan}, Hubbard ladders 
\cite{hubbardladder,vojta,liang2} and t-J ladders\cite{scalapino2}. 
Also several coupled chains at different dopings have been considered
\cite{fermion,nishimoto} as well as flux phases in these systems \cite{marston}. 
 Time reversal symmetry-broken fermionic ladders have been studied in 
\cite{time} and power laws in spinless-fermion ladders in \cite{bourbonnais}.  
Long-range Coulomb interactions in the 1D electron gas and the formation of 
a Wigner crystal was studied in \cite{ziosi}. Several phases including the Wigner 
crystal, incompressible and compressible liquid states, stripe and pairing phases, 
have been found using DMRG for 2D electrons in high magnetic fields considering 
different Landau levels\cite{landau}. Persistent currents in mesoscopic 
systems have been considered in \cite{rings}. 

Quite large quasi-2D systems can be reached, for example in
\cite{liangpang} where a 4x20 lattice was considered to study ferromagnetism 
in the infinite-U Hubbard model; the ground state of a 4-leg t-J ladder in
\cite{w1}; the one and two hole ground state in 9x9 and 10x7 t-J
lattices in \cite{w2}; a doped 3-leg t-J ladder in \cite{w3}; the study of
striped phases in \cite{arrigoni}; domain walls in 19x8 t-J systems 
in \cite{w4}; 
the 2D t-J model in \cite{bishop} and the magnetic polaron in a 9x9 t-J 
lattice in 
\cite{white5}. Also big $CaV_4O_9$ spin-1/2 lattices reaching 24x11 
sites\cite{cavo} have been studied. 
There have been some recent attempts to implement DMRG in two and higher dimensions
\cite{su,nishimoto3,nishino5,sierra3d,henelius} but the performance is still 
poorer than in 1D. A recent extension using a two-step DMRG algorithm for highly 
anisotropic spin systems has shown promising results\cite{moukouri5}.

Impurity problems have been studied for example in 
one- \cite{teo} and two-impurity \cite{egger} 
Kondo systems, in spin chains \cite{impchains} and in Luttinger Liquids 
\cite{meden}.
There have also been applications to Kondo and  Anderson lattices
\cite{kondoins,kondolatt,kondoneck,kondoxavier,kondonecklace,kondojuoza,
kondoguerrero,watanabe,ian,xavier}, 
Kondo lattices with localized 
$f^2$ configurations \cite{wata}, the two-channel Kondo lattice on a ladder 
\cite{kondomoreno}, 
a t-J chain coupled to localized Kondo
spins\cite{chen} and ferromagnetic Kondo models for manganites \cite{riera,
kondodaniel,imada}.

\section{Other extensions to DMRG}

There have been several extensions to DMRG like the inclusion of
symmetries to the method such as spin and  
parity\cite{ramasesha,affleck,simetrias}. Total spin conservation and 
continuous
symmetries have been treated in \cite{ian} and in interaction-round a
face
Hamiltonians\cite{sierrairf}, a formulation that can be applied to
rotational-invariant sytems like $S=1$ and 2 chains\cite{wada}.
A momentum representation of this technique \cite{nishimoto,xiang2,nishimoto3} 
that allows for
a diagonalization in a fixed momentum subspace has been developed
as well as  applications in dimension higher than one\cite{cavo,su,
nishimoto3,ducroo} and
Bethe lattices\cite{pastor}. 
The inclusion of symmetries is essential to the method since it allows to
consider a smaller number of states, enhance precision and obtain eigenstates
with definite quantum numbers. 
Other recent applications have been in nuclear shell model calculations
where a two level pairing model has been considered\cite{dukelsky} and
in the study of ultrasmall superconducting grains, in this case, using
the particle (hole) states around the Fermi level as the system
(environment) block\cite{duksierra}.

A very interesting and successful application is a recent work in 
High Energy Physics\cite{delgadoqcd}.
Here the DMRG is used in an asymptotically free model
with bound states, a toy model for quantum chromodynamics, namely
the two dimensional delta-function potential. For this
case an algorithm similar to the momentum space DMRG\cite{xiang2} was used
where the block and environment consist of low and high energy states
respectively. The results obtained here are much more accurate than with the
similarity renormalization group\cite{wilson2} and a generalization to
field-theoretical models is proposed based on the discreet light-cone
quantization in momentum space\cite{dlcq}.
Below we briefly mention other important extensions, leaving the calculation
of dynamical properties for the next Section.

\subsection{Classical systems}

The DMRG has been very successfully extended to study classical
systems. For a detailed description we refer the reader to Ref.
\cite{nishino}. Since 1D quantum systems are related to 2D classical
systems\cite{clasico}, it is natural to adapt DMRG to the classical 2D
case. This method is based on the renormalization group transformation
for the transfer matrix $T$ (TMRG). It is a variational method that
maximizes the partition function using a limited number of degrees of
freedom, where the variational state is written as a product of local
matrices\cite{ostlund}. 
For 2D classical systems, this algorithm is superior to the
classical Monte Carlo method in accuracy, speed and in the possibility of
treating much larger systems. A recent improvement of this method considering
periodic boundary conditions is given in \cite{gendiar2} and a detailed 
comparison between symmetric and asymmetric targetting is done in 
\cite{shibata2}. TMRG has also been successfully used to renormalize 
stochastic transfer matrices in a study of cellular automatons\cite{stochastic}.
The calculation of thermodynamical properties of 3D classical statistical systems
has been proposed\cite{nishino5} where the eigenstate of the transfer matrix with maximum 
eigenvalue is represented by the product of local tensors optimized 
using DMRG.

A further improvement to this method  is based on the corner transfer
matrix\cite{baxter}, the CTMRG\cite{okunishi,takasaki,nishino3,ritter} and can 
be generalized to any dimension\cite{okunishi2}.

It was first applied to the Ising model\cite{nishino,drz,drz2,kaulke} 
and also to the
Potts model\cite{carlon}, where very accurate density profiles and
critical indices were calculated. Further applications have included
non-hermitian problems in equilibrium and non-equilibrium
physics.
In the first case, transfer matrices may be non-hermitian and several
situations have been considered: a model for the Quantum Hall
effect\cite{kondev}, the $q$-symmetric Heisenberg chain
related to the conformal series of critical models\cite{peschel} and the 
anisotropic triangular nearest and next-nearest neighbour Ising 
models\cite{gendiar3}.
In the second case, the adaptation of the DMRG to non-equilibrium physics 
like the asymmetric exclusion problem\cite{hieida2} and
reaction-diffusion problems \cite{peschel1,carlon1} has shown to be very
successful. It has also been applied to stochastic lattice models like in 
\cite{jef} and to the 2D XY model \cite{chung3}.

\subsection{Finite-temperature DMRG}

The adaptation of the DMRG method for classical systems
paved the way for the study of 1D quantum systems at non zero
temperature, by using the Trotter-Suzuki
method \cite{bursill2,trotter,xiangwang,ueda2,shibata}. In this
case the system is infinite and the finiteness is in the level of the
Trotter approximation. Standard DMRG usually
produces its best results for the ground state energy and less accurate
results for higher excitations. A different situation occurs here: the
lower the temperature, the less accurate the results. 

Very nice results have been obtained for the dimerized, $S=1/2$, XY model,
where the specific heat was calculated involving an extremely small basis
set\cite{bursill2} ($m=16$), the agreement with the exact solution being
much better in the case where the system has a substantial gap. 
It has also been used to calculate thermodynamical properties of the
a\-ni\-so\-tropic $S=1/2$ Heisenberg model, with relative errors for the spin
susceptibility of less than $10^{-3}$ down to temperatures of the order of
$0.01J$ keeping $m=80$ states\cite{xiangwang}. A complete study of
thermodynamical properties like magnetization, susceptibility, specific
heat and temperature dependent correlation functions  for the 
$S=1/2$ and 3/2 Heisenberg models was done in \cite{xiangt}.
Other applications have been the
calculation of the temperature dependence of the charge and spin gap in
the Kondo insulator\cite{ammon1}, the calculation of thermodynamical
properties of ferrimagnetic chains\cite{scholl} and spin ladders\cite{wanglu},
the study of impurity
properties in spin chains\cite{rommer,maruyama}, frustrated quantum spin
chains\cite{maisinger}, t-J\cite{ammon} and spin ladders\cite{naefwang} 
and dimerized frustrated
Heisenberg chains\cite{klumper}.

An alternative way of incorporating temperature into the DMRG procedure
was developed by Moukouri and Caron\cite{moukouri}. They considered the
standard DMRG taking into account several low-lying target states (see
Eq.~\ref{eq:pl}) to
construct the density matrix, weighted with the Boltzmann factor 
($\beta$ is the inverse temperature):
\begin{equation}
\label{eq:pl2}   
\rho_{ii'}=\sum_l e^{-\beta E_l} \sum_j \phi_{l,ij} \phi_{l,i'j}
\end{equation}
With this method they performed reliable calculations of the magnetic
susceptibility of quantum spin chains with $S=1/2$ and $3/2$, showing
excellent agreement with Bethe Ansatz exact results. They
also calculated low temperature thermodynamical properties of the 1D Kondo
Lattice Model\cite{moukouri3} and of organic conductors \cite{moukouri4}. 
Zhang et al.\cite{zhang} applied the
same method in the study of a magnetic impurity embedded in a quantum spin
chain.

\subsection{Phonons, bosons and disorder}

A significant limitation to the DMRG method is that it requires a finite
basis and calculations in problems with infinite degrees of freedom per
site require a large truncation of the basis states\cite{moukouri2}.
However, Jeckelmann and White developed a way of
including phonons in DMRG calculations by transforming each boson site
into several artificial interacting two-state pseudo-sites and then
applying DMRG to this interacting system\cite{jeck} (called the
``pseudo-site system"). The idea is based on the fact that DMRG is much
better able to handle several few-states sites than few many-state
sites\cite{noackberlin}. The key idea is to substitute each boson site
with $2^N$ states 
into $N$ pseudo-sites with 2 states\cite{jeckbook}. They
applied this method to the Holstein model for several hundred sites
(keeping more than a hundred states per phonon mode) obtaining negligible
error. In addition, up to date, this method is the most accurate one to
determine the ground state energy of the polaron problem (Holstein model
with a single electron).

An alternative method (the ``Optimal phonon basis")\cite{jeck2} is a
procedure for generating a controlled truncation of a large Hilbert
space, which allows the use of a very small optimal basis without
significant loss of accuracy. The system here consists of
only one site and the environment has several sites, both having
electronic and phononic degrees of freedom.
The density matrix is used to trace out the degrees of freedom of the
environment and extract the most relevant states of the site in question.
In following steps, more bare phonons are included to the optimal basis
obtained in this way. This method was successfully applied to study the 
interactions induced by quantum fluctuations in quantum strings, as an application 
to cuprate stripes\cite{nishiyama3} and the dissipative two-state system
\cite{nishiyama5}.
 A variant of this scheme is the ``four block
method", as described in \cite{bursillphonon}. They obtain very accurately
the Luttinger liquid-CDW insulator transition in the 1D Holstein model for
spinless fermions.
 
The method has also been applied to pure bosonic systems such as the
disordered bosonic Hubbard model\cite{krish}, where gaps, correlation
functions and superfluid density are obtained. The phase diagram for the
non-disordered Bose-Hubbard model, showing a reentrance of the superfluid
phase into the insulating phase was calculated in Ref. \cite{monien}. 
It has also been used to study a chain of oscillators with optical phonon 
spectrum \cite{chung} and optical phonons in the quarter-filled Hubbard model for 
organic conductors\cite{maurel1}. 

The DMRG has also been generalized to 1D random and disordered systems, and 
applied
to the random antiferromagnetic and ferromagnetic
Heisenberg  chains\cite{hida}, including quasiperiodic exchange  
modulation\cite{hida1} and a detailed study of the Haldane\cite{hida2}
and Griffiths phase\cite{griffiths} in these systems.
Strongly disordered spin ladders have been considered in
\cite{spinladders}. 
It has also been used in disordered Fermi systems such as the spinless
model\cite{schmitt,schmitt2}. In particular, the transition from the Fermi glass to
the Mott insulator and the strong enhancement of persistent currents in the
transition was studied in correlated one-dimensional disordered
rings\cite{jala}. Disorder-induced crossover effects at quantum critical 
points were studied in \cite{carlonigloi}.

\subsection{Molecules and Quantum Chemistry}

There have been several applications to molecules and polymers, such as
the Pariser-Parr-Pople (PPP) Hamiltonian for a cyclic polyene\cite{ppp}
(where long-range interactions are included), magnetic Keplerate molecules
\cite{keplerate}, molecular Iron rings\cite{bruce} 
 and polyacenes considering long range interactions 
\cite{polyacenes}. 
It has also been applied to conjugated organic systems (polymers), adapting 
the DMRG to take into account the most important symmetries in 
order to obtain the desired excited states\cite{ramasesha}. Also conjugated 
one-dimensional semiconductors \cite{barford} have been studied, in which
the standard approach can be extended to complex 1D oligomers where the 
fundamental repeat is not just one or two atoms, but a complex molecular 
building block. Relatively new fields of application are the calculation of 
dynamical properties in the Rubinstein-Duke model for reptons \cite{reptons} 
and excitons in dendrimer molecules\cite{dendrimers}.

Recent attempts to apply DMRG to the {\it ab initio} calculation of
electronic states in molecules have been successful\cite{whitemol,whiteorth,
legezaqc,daulwhite}. 
Here, DMRG is applied within the conventional quantum
chemical framework of a finite basis set with non-orthogonal basis
functions centered on each atom. After the standard Hartree-Fock (HF)
calculation in which a Hamiltonian is produced within the orthogonal HF
basis, DMRG is used to include correlations beyond HF, where each orbital
is treated as a ``site" in a 1D lattice. One important difference with
standard DMRG is that, as the interactions are long-ranged, several
operators must be kept, making the calculation somewhat cumbersome.
However, very accurate results have been obtained in a
check performed in a water molecule (keeping up to 25 orbitals and
$m\simeq 200$ states per block), obtaining an offset of 0.00024Hartrees
with respect to the exact ground state energy\cite{bau}, a better
performance than any other approximate method\cite{whitemol}.

In order to avoid the non-locality introduced in the treatment explained
above, White introduced the concept of {\it orthlets}, local, orthogonal
and compact wave functions that allow prior knowledge about singularities
to be incorporated into the basis and an adequate resolution for the
cores\cite{whiteorth}. The most relevant functions in this basis are
chosen via the density matrix. An application based on the combination with 
the momentum version of DMRG is used in \cite{momentum} to calculate the 
ground state of several molecules.

\section{Dynamical correlation functions}

The DMRG was originally developed to calculate static ground state
properties  and low-lying energies. However, it can also be useful to  
calculate dynamical response functions. These are of great interest in  
condensed matter physics in connection with experiments such as nuclear  
magnetic resonance (NMR), neutron scattering, optical absorption,  
photoemission, etc. We will describe three different methods in this Section.

A recent development for calculating response functions in single 
impurity systems in the presence of a magnetic field was done in 
\cite{hofstetter} by using the DMRG within Wilson's NRG to obtain the 
Green's function. 

An interesting extension of DMRG to tackle time-dependent quantum many-body systems 
out of equilibrium was considered by Cazalilla and Marston\cite{cazalilla}.

\subsection{Lanczos and correction vector techniques}

An effective way of extending the basic ideas of this method
to the calculation of dynamical quantities is described in Ref.\cite{karendin}.
It is important to notice here that due to the particular real-space
construction, it is not possible to fix the momentum as a quantum number.  
However, we will show that by keeping the appropriate target states, a
good  value of momentum can be obtained.

	We want to calculate the following dynamical correlation function
at $T=0$:
\begin{equation}
\label{eq:ca}
C_A(t-t')=\langle\psi_0|A^{\dagger}(t) A(t')|\psi_0 \rangle ,
\end{equation}
where $A^{\dagger}$ is the Hermitean conjugate of the operator $A$, $A(t)$ is
the Heisenberg representation of $A$, and $|\psi_0 \rangle $ is the ground
state of the system. Its Fourier transform is:
\begin{equation}
C_A(\omega )=\sum_n |\langle \psi_n | A |\psi_0 \rangle |^2 \; 
\delta (\omega - (E_n-E_0)),
\end{equation}
where the summation is taken over all the eigenstates $|\psi_n \rangle$ of
the Hamiltonian $H$ with energy $E_n$, and $E_0$ is the ground state energy.

Defining the Green's function
\begin{equation}
\label{eq:din}
G_A(z)=\langle \psi_0 | A^{\dagger}(z-H)^{-1} A |\psi_0 \rangle,
\end{equation}
the correlation function $C_A(\omega)$ can be obtained as
\begin{equation}
C_A(\omega)=-\frac{1}{\pi}\lim_{\eta\to 0^+}{\rm Im} \; G_A(\omega+i\eta +E_0).
\end{equation}

The function $G_A$ can be written in the form of a continued fraction:
\begin{equation}
\label{eq:frac}
G_A(z)=\frac{\langle \psi_0 | A^{\dagger} A|\psi_0\rangle}{z-a_0-\frac{b_1^2}
{z-a_1-\frac{b_2^2}{z-...}}}
\end{equation}	
The coefficients $a_n$ and $b_n$ can be obtained using the following
recursion equations \cite{carlos,proj}: 
\begin{equation}
|f_{n+1}\rangle =H|f_n\rangle -a_n|f_n\rangle -b_n^2|f_{n-1}\rangle
\end{equation}
where
\begin{eqnarray}
|f_0\rangle &=& A|\psi_0\rangle \nonumber \\
a_n&=&\langle f_n|H|f_n\rangle/\langle f_n|f_n\rangle, \nonumber \\
b_n^2&=&\langle f_n|f_n\rangle/\langle f_{n-1}|f_{n-1}\rangle; \;\; b_0=0
\end{eqnarray}

For finite systems the Green's function $G_A(z)$ has a finite number of
poles so only a certain number of coefficients $a_n$ and $b_n$ have to be
calculated. The DMRG technique presents a good framework to calculate such
quantities. With it, the ground state, Hamiltonian and the operator $A$
required for the evaluation of $C_A(\omega)$ are obtained. An important
requirement is that the reduced Hilbert space should also describe with great
precision the relevant excited states $|\psi_n \rangle $. This is achieved by
choosing the appropriate target states. 
For most systems
it is enough to consider as target states the ground state $|\psi_0\rangle$ and
the first few $|f_n\rangle $ with $n=0,1...$ and $|f_0\rangle=
A|\psi_0\rangle$ as described above. In doing so,  states
in the reduced Hilbert space
relevant to the excited states connected to the ground state via the
operator of interest $A$ are included. The fact that  $|f_0\rangle$ is
an excellent trial state, in particular, for the lowest triplet excitations 
of the two-dimensional antiferromagnet was shown in Ref.~\cite{linden}.
Of course, if the number $m$ of states kept per block is fixed, the more 
target states  considered, the less precisely each one of them is
described. An optimal number of target states
and $m$ have to be found for each case. Due to this reduction, the
algorithm can be applied up to certain lengths, depending on the states
involved. For longer chains, the higher energy excitations will become
inaccurate. Proper sum rules have to be calculated to determine the errors
in each case.

	As an application of the method we calculate
\begin{equation}
\label{eq:szzn}
S^{zz}(q,\omega)=\sum_n |\langle \psi_n | S^z_q |\psi_0 \rangle |^2 \; 
\delta (\omega - (E_n-E_0)),
\end{equation}
for the 1D isotropic Heisenberg model with spin $S=1/2$. 

The spin dynamics of this model has been extensively studied. The lowest
excited states in the thermodynamic limit 
are the  des~Cloiseaux-Pearson triplets 
\cite{descloi}, having total spin $S^T=1$. The dispersion of this
spin-wave branch is $\omega^l_q=\frac{J\pi}{2}|\sin (q)|$.
Above this lower boundary there exists a two-parameter continuum of
excited triplet states that have been calculated using the Bethe Ansatz
approach \cite{yam} with an upper boundary given by 
$\omega^u_q=J\pi|\sin(q/2)|$.
It has been shown \cite{1haas}, however, 
that there are excitations above this upper
boundary due to higher order scattering processes,
with a weight that is at least one order of magnitude lower
than the spin-wave continuum.

In Fig. 2 we show the spectrum for $q=\pi$
and $N=24$ for different values of $m$, where exact results are available for 
comparison. The delta
peaks of Eq.~(\ref{eq:szzn}) are broadened by a Lorentzian for
visualizing purposes. As
expected, increasing $m$ gives more precise results for the higher
excitations. This spectra has been obtained using the infinite-system method 
and more precise results are expected using the finite-system method, as 
described later.

\begin{figure}[htbp]
\begin{center}
\vspace*{0.5cm}
\epsfxsize=3.3in
\epsfysize=2.75in
\epsffile{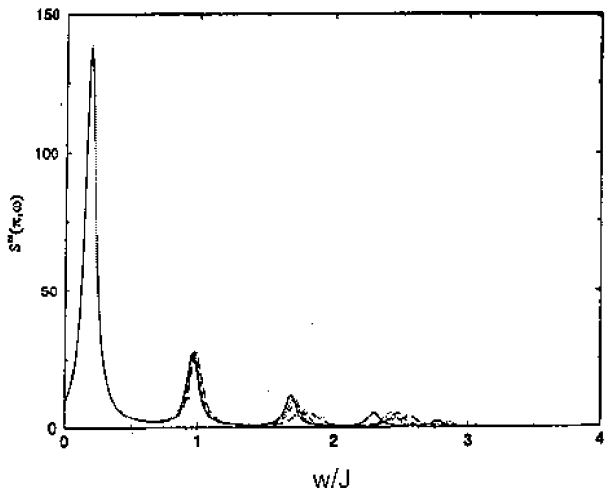}
\end{center}
\vspace*{0.3cm}
\caption[]{Spectral function for a Heisenberg chain with $N=24$ and
$q=\pi$. Full line: exact result {\protect{\cite{haas}}}. The rest are
calculated using DMRG with $m=100$ (long-dashed line), $m=150$ (dashed
line) and $m=200$ (dotted line). }
\label{fig1karen}
\end{figure}

In Fig. 3 we show the spectrum for two systems lengths and $q=\pi$
and $q=\pi/2$ keeping $m=200$ states and periodic boundary conditions.
For this case it was enough to take 3 target states,
{\it i. e.~} $|\psi_0\rangle$, $|f_0\rangle = S^z_{\pi}|\psi_0\rangle$ and
$|f_1\rangle$.
Here we have used $\sim 40$ pairs of coefficients $a_n$ and $b_n$, but
we noticed that if we considered only the first ($\sim 10$) coefficients
$a_n$ and $b_n$, the spectrum at low energies remains essentially unchanged.
Minor differences arise at $\omega /J\simeq 2$. This is another indication
that only the first $|f_n\rangle$ are relevant for the low energy
dynamical properties for finite systems.

In the inset of
Fig. 3 the spectrum for $q=\pi/2$ and $N=28$ is shown. For this case
we considered 5 target states {\it i. e.~} $|\psi_0\rangle$, 
$|f_0\rangle = S^z_{\pi/2}|\psi_0\rangle$, $|f_n\rangle\; n=1,3$ and
$m=200$. Here, and for all the cases considered, we have verified that
the results are very weakly dependent on the weights $p_l$ of the target
states (see Eq.(\ref{eq:pl}))   
as long as the appropriate target states are chosen. 
For lengths where this value of $q$ is not defined we took the nearest
value.


\begin{figure}[htbp]   
\begin{center}
\epsfxsize=3.3in
\epsfysize=2.75in   
\epsffile{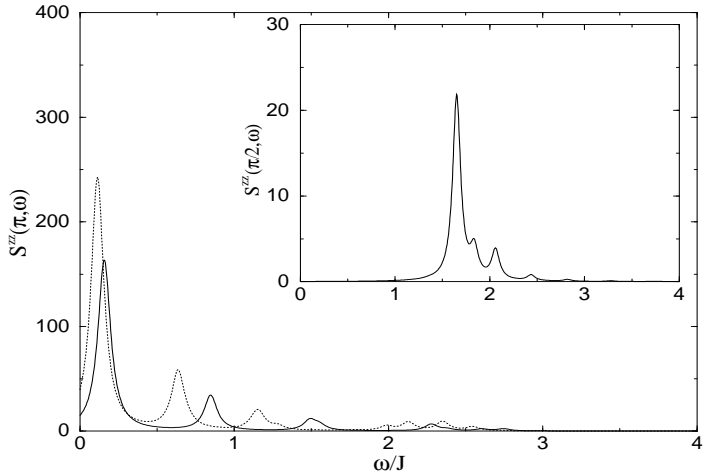}
\end{center}                          
\caption[]{Spectral densities for $q=\pi$, $N=28$ (continuous line) and 
$N=40$ (dotted line). Inset: Spectral density
for  $q=\pi/2$ for $N=28$ ($\eta=0.05$). }
\label{fig2karen}
\end{figure}

	Even though we are including states with a given momentum
as target states, due to the particular real-space construction of the
reduced Hilbert space, this translational symmetry is not fulfilled
and the momentum is not fixed. To check how the reduction on the Hilbert
space influences the momentum $q$ of the target state 
$|f_0\rangle =S^z_q|\psi_0\rangle$, we
calculated the expectation values 
$\langle \psi_0 |S_{-q'}^z S_q^z|\psi_0\rangle$ 
for all $q'$. If the momenta of the states were well defined, this value is
proportional to $\delta_{q-q'}$ if $q\neq 0$. For $q=0$, $\sum_r S^z_r=0$.

The momentum distribution for $q=\pi$ is shown in Fig. 4 in a
semilogarithmic scale where the $y$-axis has been shifted  by
.003 so as to have well-defined logarithms. 
We can see here that the momentum is better
defined, even for much larger systems, but, as expected, more weight on
other $q'$ values arises for larger $N$.


\begin{figure}[htbp]
\begin{center}
\epsfig{file=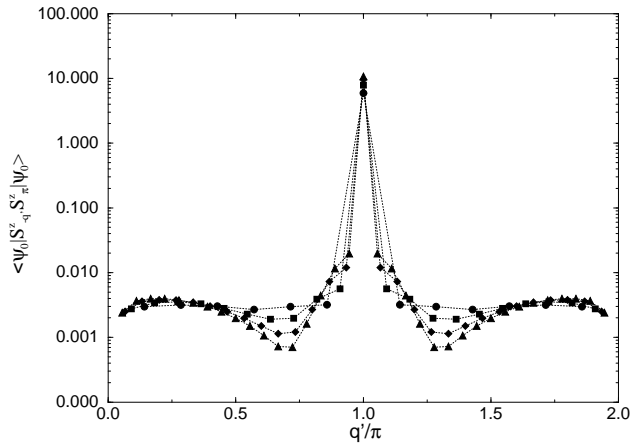,width=6cm,angle=-90}
\end{center}
\vspace*{0.5cm}
\caption[]{Momentum weights  of a target
state with $q=\pi$ for $N=28$ (circles),  $N=44$ (squares), $N=60$
(diamonds) and $N=72$ (triangles). The dotted lines are a guide to the
eye. }
\label{fig3karen}
\end{figure}
	As a check of the approximation we calculated the sum rule
\begin{equation}
\frac{1}{4\pi^2}\int_0^{\infty}d\omega \int_{q=0}^{2\pi}
S^{zz}(q,\omega)\equiv \langle \psi_0 |(S_{r=0}^z)^2|\psi_0\rangle
=\frac{1}{4}
\end{equation}
for $N=28$, 5 target states and $m=200$. We obtain a relative error of 
0.86\%. 

Recently, important improvements to this method have been published
\cite{kuhner}: By considering the finite system method in
open chains, K\"uhner and White obtained a higher precision in dynamical
responses of spin chains. In order to define a momentum in an open chain
and to avoid end effects, they introduce a filter function with
weight centered in the middle of the chain and zero at the boundaries.

Recent applications of this method include the calculation of excitations in
spin-orbital models (SU(4)) in a magnetic field\cite{haas2}, spin dynamics in
models for cuprate spin ladders including cyclic spin exchange\cite{nunner},
optical conductivity of the ionic Hubbard model\cite{brune1}, excitations in 
the one-dimensional Bose-Hubbard model\cite{monien2} and the optical response in 
1D Mott insulators\cite{bolech}.

In this section we have presented a method of calculating dynamical
responses with DMRG. Although the basis truncation is big, 
this method keeps only the most relevant states and, for example,
even by considering a $0.1\%$ of the
total Hilbert space (for $N=28$ only $\sim$ 40000 states are kept)
a reasonable description of the low energy excitations is obtained. 
We show that it is also possible to obtain
states with well defined momenta if the appropriate target states are used.

\subsubsection{Correction vector technique}

Introduced in Ref.~\cite{cvramasesha} in the DMRG context and improved in 
Ref.~\cite{kuhner}, this method 
focuses on a particular energy or energy window, allowing a more precise 
description in that range and the possibility of calculating spectra for
higher energies. Instead of using the tridiagonalization of the
Hamiltonian, but in a similar spirit regarding the important target
states to be kept, the spectrum can be calculated for a given $z=w+i\eta$
by using a correction vector (related to the operator $A$ that can
depend on momentum $q$). 

Following (\ref{eq:din}), the (complex) correction vector
$|x(z)\rangle$ can be defined as: 
\begin{equation}
|x(z)\rangle = \frac{1}{z-H}A |\psi_0 \rangle
\end{equation}
 so the Green's function can be calculated as 
\begin{equation}
G(z)=\langle \psi_0 |A^{\dagger} |x(z)\rangle 
\end{equation}

Separating the correction vector in real and imaginary parts 
$|x(z)\rangle = |x^r(z)\rangle + i |x^i(z)\rangle$ we obtain
\begin{equation}
((H-w)^2 + \eta^2)|x^i(z)\rangle = -\eta A |\psi_0 \rangle 
\end{equation}
 and
\begin{equation}
|x^r(z)\rangle= \frac{1}{\eta}(w-H)|x^i(z)\rangle 
\end{equation}
The former equation is solved using the conjugate gradient method.
In order to keep the information of the excitations at this particular
energy the following states are targeted in the DMRG iterations: The
ground state $|\psi_0 \rangle$, the first Lanczos vector $A |\psi_0
\rangle$ and the correction vector $|x(z)\rangle$.
Even though only a certain energy is focused on, DMRG gives the correct
excitations for an energy range surrounding this particular point so that
by running several times for nearby frequencies, an approximate spectrum
can be obtained for a wider region \cite{kuhner}.

A variational formulation of the correction vector technique which leads to
more accurate excited energies and spectral weights has been developed in
\cite{ddmrg}. It has been successfully applied to calculate the optical 
conductivity of Mott insulators\cite{florian}.

\subsection{Moment expansion}

This method\cite{pang} relies on a moment expansion of the dynamical
correlations
using sum rules that depend only on static correlation functions which can
be calculated with DMRG. With these moments, the Green's functions can be
calculated using the maximum entropy method.

The first step is the calculation of sum rules. As an example, and
following \cite{pang}, the spin-spin correlation
function $S^z(q,w)$ of the Heisenberg model is calculated
where the operator $A$ of
Eq.~(\ref{eq:ca}) is $S^z(q)=N^{-1/2}\sum S^z(l) \exp(iql)$
 and the sum rules are\cite{hohenberg}:
\begin{eqnarray}
m_1(q)&=&\int_0^\infty \frac{dw}{\pi}\frac{S^z(q,w)}{w}=
\frac{1}{2}\chi(q,w=0) \nonumber \\
m_2(q)&=&\int_0^\infty \frac{dw}{\pi} w \frac{S^z(q,w)}{w}=
\frac{1}{2}S^z(q,t=0) \nonumber \\
m_3(q)&=&\int_0^\infty \frac{dw}{\pi} w^2 \frac{S^z(q,w)}{w}=
-\frac{1}{2}\langle [[H,S^z(q)],S^z(-q)] \rangle \nonumber \\
&=& 2[1-\cos(q)]\sum_i\langle S^+_{i}S^-_{i+1}+S^-_{i}S^+_{i+1}\rangle
\end{eqnarray}
where $\chi(q,w=0)$ is the static susceptibility. These sum rules can be
easily generalized to higher moments:
\begin{eqnarray}
m_l(q)&=&\int_0^\infty \frac{dw}{\pi} w^{l-1} \frac{S^z(q,w)}{w} \nonumber
\\
&=& -\frac{1}{2} \langle [[H,...,[H,S^z(q)]...],S^z(-q)]\rangle 
\end{eqnarray}
for $l$ odd. 
A similar expression is obtained for $l$ even, where the outer square
bracket is replaced by an anticommutator and the total sign is changed.
Here $H$ appears in the commutator $l-2$ times.

Apart from the first moment which is given by the static susceptibility,
all the other moments can be expressed as equal time correlations (using a
symbolic manipulator). The static susceptibility $\chi$ is calculated by
applying
a small field $h_q\sum_i n_i \cos(qi)$ and calculating the density
response $\langle n_q \rangle = 1/N \sum_i \langle n_q \rangle \cos(qi)$
with DMRG. Then $\chi= \langle n_q \rangle / h_q$ for $h_q\to 0$.
These moments are calculated for several chain lengths and extrapolated to
the infinite system. 
Once the moments are calculated, the final spectra is constructed via the
Maximum Entropy method (ME), which has become a standard way to extract
maximum information from incomplete data (for details see Ref. \cite{pang}
and references therein). Reasonable spectra are obtained for the XY and
isotropic models, although information about the exact position of the
gaps has to be included. Otherwise, the spectra are only qualitatively
correct.
This method requires the calculation of a large amount of moments in order
to get good results: The more information given to the ME equations, the
better the result.

\subsection{Finite temperature dynamics}

In order to include temperature in the calculation of dynamical quantities, 
the Transfer Matrix RG described above
(TMRG\cite{bursill2,xiangwang,shibata})
was extended to obtain imaginary time correlation 
functions\cite{wangbook,mutou,naef}. 
After Fourier transformation in the imaginary time axis, analytic
continuation from imaginary to real frequencies is done using maximum
entropy (ME). The combination of the TMRG and ME is free from statistical
errors and the negative sign problem of Monte Carlo methods. 
Since we are dealing with the transfer matrix, the
thermodynamic limit can be discussed directly without extrapolations.
However, in the present scheme, only local quantities can be calculated.

A systematic investigation of local spectral functions is done in Ref.
\cite{naef} for the anisotropic Heisenberg antiferromagnetic chain. The
authors 
obtain good qualitative  results especially for high temperatures but a  
quantitative description of peaks and gaps are beyond the method, due to
the severe intrinsic limitation of the analytic continuation.
This method was also applied with great success to the 1D Kondo
insulator\cite{mutou}. The temperature dependence of the
local density of states and local dynamic spin and charge correlation
functions were calculated. 

\section{Conclusions}

We have presented here a very brief description of the Density Matrix
Renormalization Group technique, its applications and extensions. The aim
of this article is to give the unexperienced reader an idea of the
possibilities and scope of this powerful, though relatively simple
method. The experienced reader can find here an extensive (however
incomplete) list of references covering most applications to
date using DMRG in a great variety of fields such as Condensed Matter,
Statistical Mechanics and High Energy Physics.

\section*{Acknowledgments}

The author acknowledges hospitality at the Centre de Recherches
Mathematiques, University of Montreal and at the Physics Department of the 
University of Buenos Aires, Argentina, where this work has been performed.
We thank S. White for a critical reading of the manuscript and
all those authors that have updated references and sent instructive comments.
K. H. is a fellow of CONICET, Argentina. Grants: PICT 03-00121-02153 and 
PICT 03-00000-00651.



\end{document}